\documentclass{emulateapj}
\usepackage{apjfonts}
\usepackage{mathptmx}
\usepackage{epsf}

\submitted{submitted to ApJ Letters}
\shorttitle{Beyond Reionization}
\shortauthors{Barton et al.}

\newcommand{\mewe}  {{\rm EW(Ly\alpha)_{\rm em}}}

\begin{document}

\title{Searching for Star Formation Beyond Reionization}

\author{\sc Elizabeth J. Barton\altaffilmark{1,2},
            Romeel Dav\'{e}\altaffilmark{1,2},
            John-David T. Smith\altaffilmark{2},
            Casey Papovich\altaffilmark{2},
            Lars Hernquist\altaffilmark{3}, and
            Volker Springel\altaffilmark{4}
}

\altaffiltext{1}{Hubble fellow}
\altaffiltext{2}{University of Arizona, Steward Observatory, 933 N.
Cherry Ave., Tucson, AZ 85721}
\altaffiltext{3}{Harvard-Smithsonian Center for Astrophysics,
60 Garden Street, Cambridge, MA 02138}
\altaffiltext{4}{Max-Planck-Institut f\"{u}r Astrophysik,
karl-Schwarzschild-Stra$\beta$e 1,
85740 Garching bei M\"{u}nchen, Germany}

\begin{abstract}

The goal of searching back in cosmic time to find star formation
during the epoch of reionization will soon be within reach.  We assess
the detectability of high-redshift galaxies by combining cosmological
hydrodynamic simulations of galaxy formation, stellar evolution models
appropriate for the first generations of stars, and estimates of the
efficiency for Ly$\alpha$ to escape from forming galaxies into
the intergalactic medium.  Our simulated observations show
that Ly$\alpha$ emission at $z \sim 8$ may be observable in the
near-infrared with 8-meter class telescopes and present-day
technology.  Not only is the detection of early star-forming objects
vital to understanding the underlying cause of the reionization of the
universe, but the timely discovery of a $z > 7$ star-forming
population --- or even an interesting upper limit on the emergent flux
from these objects --- will have implications for the design of the
next generation of ground- and space-based facilities.

\end{abstract}


\section{Introduction}

Recent observations have significantly advanced our understanding of
the reionization of the universe.  Estimates of the Thomson optical
depth from the {\it Wilkinson Microwave Anisotropy Probe (WMAP)}
suggest that the reionization epoch began at redshifts $14 \lesssim z
\lesssim 20$ \citep{WMAP1,WMAP2} and Gunn-Peterson troughs in distant
quasars indicate that reionization ended at $z \sim 6$
\citep{Bec01,Fan02}.  However, the known population of quasars cannot
produce sufficient ionizing radiation \citep{Fan01}, so star-forming
galaxies are most likely the dominant source of ionizing photons at
redshifts $z > 7$, a view supported by theoretical studies of
reionization \citep[e.g.,][]{Tin73,Sok03a}.  The direct detection of these
objects lies at the next frontier in the study of the evolution of the
early universe.

Discoveries of star-forming objects are progressing to ever higher
redshifts, with successful results from both broad-band color selection
\citep[e.g.,][]{Yan03,Sta03,Bun03,Dic03} and Ly$\alpha$ emission line
searches \citep[e.g.,][]{Rho03,Mai03,Cub03}.  Although the number counts
of $I$-band ``dropouts'' (at $5.5 \lesssim z \lesssim 6.5$) are still
heavily debated, it is clear that a substantial population of galaxies
with detectable continuum exists at these redshifts.  The next step
--- discovering a population of objects before the
end of reionization --- lies behind a technological barrier:  most
of the light from $z > 7$ objects falls in the near-infrared, where
backgrounds provide significant challenges from the ground.
High spectral resolution observations provide some relief; dispersing the
light eliminates much of the background noise and allows the detection
of line emission \citep[e.g.,][]{Tho95,Tho96,Pan03}.

At first glance, weak star formation and a neutral intergalactic
medium (IGM) would appear to critically hamper the detection of
Ly$\alpha$ at high redshift.  However, a top-heavy initial stellar
mass function (IMF) and low metallicity may boost the Ly$\alpha$ flux
of the earliest stars \citep[e.g.,][]{Bro01,Sch03}.  In addition,
this emission may penetrate even a neutral IGM with reasonable
efficiency \citep{Hai02,San03}.

Taken together, the arguments above strongly motivate the search for
Ly$\alpha$ emission at redshifts $z \geq 7$ using narrow-band imaging in
the near-infrared.  Here, we describe a search for Ly$\alpha$ in a narrow
$J$-band ``window'' in the sky background, corresponding to $z\sim 8$.

\section{Star-Forming Objects at $z \approx 8$}

\subsection{Hydrodynamical Simulations of Early Star Formation}

We examine the star formation rates in the cosmological
simulations of \citet{Spr03a}.  Employing a novel treatment of smoothed
particle hydrodynamics and a multiphase description of
star-forming gas to incorporate feedback and stellar winds 
\citep{Spr02,Spr03b}, \citet{Spr03a} obtained a
numerically converged prediction for the cosmic star formation rate (SFR)
as a function of redshift that agrees
with the observed star formation history at low redshift.\footnote{We 
note an error in figure 12 of \citet{Spr03a} in which the observational
estimates of the SFR were plotted too high by a factor of
$h^{-1} = 1.4$.  When corrected, the observed points are in better agreement
with their theoretical estimates.}
Here, we focus on the $10\ h^{-1}\ {\rm comoving\
Mpc}$ Q5 simulation of a {\it WMAP}-concordant cosmology with
$2\times324^3$ particles.  The more luminous sources in the
simulations have star formation rates in the range of ${\rm 10^{-2}
\leq \dot{M} \lesssim 1}$~M$_{\sun}$/yr.  With typical dynamical
masses of ${\rm 10^8-10^9\ M_{\sun}}$ at $z=8$, their intrinsic
circular velocities are generally ${\rm 30-50\ km\ s^{-1}}$.  The
simulations predict an early rise in star formation activity at $z
\geq 5$; at $z=8$, the star formation rate density is within a factor
of $\sim$2 of its peak, relatively independent of the details of
the physics of star formation and feedback (Hernquist \& Springel 2003).

\subsection{Ly$\alpha$ Emission and Absorption: the IMF, the IGM, and the ISM}\label{sec: theory}

The amount of Ly$\alpha$ that we can observe from young star-forming objects
depends on the IMF and metallicity in early galaxies, the amount of
attenuation within the galaxy itself, and the amount of scattering in
the intergalactic medium:
\begin{equation}
{\rm F_{Ly\alpha,obs}=e^{-\tau_{dust}}\ (1-f_{esc})\ f_{IGM}\ F_{Ly\alpha,em} }
\label{eqn:lya}
\end{equation}
where ${\rm F_{Ly\alpha,obs}}$ is the observable Ly$\alpha$ flux per
unit of star formation, {\rm $\tau_{\rm dust}$} is the optical depth
due to dust in the galaxy, ${\rm f_{esc}}$ is the fraction of ionizing
photons that escape the star-forming galaxy and thus create no
Ly$\alpha$ photons at the source, and ${\rm f_{IGM}}$ is the total
fraction of Ly$\alpha$ photons that penetrate the intergalactic medium
and reach us.  ${\rm F_{Ly\alpha,em}}$ is the maximum,
ionization-bounded flux of Ly$\alpha$ photons produced per unit of
star formation; it depends on the IMF and metallicity.  Recent models
and the interpretation of high-redshift galaxies provide some
constraints on these quantities.

{\bf The IMF:} Theoretical models predict a top-heavy IMF for the
first generation of stars formed in metal-free gas
\citep{Bro99,Bro02,Abe00}.  These stars can radiate more than an
order of magnitude more ionizing photons per unit solar mass than
a Salpeter IMF, resulting in much stronger Ly$\alpha$ and
He\,{\sc II}($\lambda1640$) \citep{Bro01,Sch03}.  However, only the initial
generations of stars will be formed from completely pristine gas;
thus, one major source of uncertainty is whether luminous bursts of
star formation later in the process ($z=8$) will be too enriched to
exhibit extremely top-heavy IMFs.

{\bf The ISM:} The factor ${\rm e^{-\tau_{dust}}\ (1-f_{esc})}$ is the
fraction of emitted ionizing photons available for conversion to
Ly$\alpha$; it includes the absorption of ionizing and Ly$\alpha$
photons from dust near the newly forming stars.  There are, at
present, few direct observational constraints on this factor at any
redshift.  At $z \sim 3$, \citet{Pet98} estimate ${\rm e}^{-\tau_{\rm dust}}
\sim 0.16 - 0.4$.  But the stellar populations at even higher redshift
are probably younger, less chemically evolved, and hence less dusty.
For the escape fraction, \citet{Ste01} combine $z \sim 3$ Lyman-break
galaxy spectra to estimate an {\it average} ${\rm f_{esc} \gtrsim
0.07-0.1}$.

{\bf The IGM:} The discovery of Gunn-Peterson troughs in the spectra
of quasars at $z \gtrsim 6$ indicate that the average neutral fraction
of the intergalactic medium exceeds $x_{\rm HI} \sim 10^{-2}$ at $z
\gtrsim 6$ \citep{Bec01,Fan02}.  At face value, this result suggests
that Ly$\alpha$ emission from beyond this redshift will not penetrate
the IGM.  However, \citet{Hai02} shows that a substantial fraction of
the emitted Ly$\alpha$ photons escape scattering by the IGM when a
galaxy ionizes a local bubble in its immediate surroundings.
Extending this idea using dynamical models of the IGM and galactic
winds, \citet{San03} derives values of f$_{\rm IGM}$ ranging from
$\lesssim 0.002$ to $\sim 1$.  This broad range illustrates the
substantial uncertainties involved in estimating the expected
Ly$\alpha$ flux. In general, penetration through the IGM improves for
galaxies with lower redshifts, higher star-formation rates, and older
(longer-duration) bursts of star formation, higher ${\rm f_{esc}}$,
outflows in which the centroid of Ly$\alpha$ is shifted redward, and
with higher ionizing backgrounds in the universe.  Because unabsorbed
ionizing photons either create Ly$\alpha$ in the source or escape to
ionize the IGM, but not both, the total emergent Ly$\alpha$ decreases
for both very high and very low escape fractions.  The fiducial models
of \citet{San03} show a broad peak in detectable Ly$\alpha$ at ${\rm
f_{esc}} \sim 0.1 - 0.8$.

\begin{figure}
\plotone{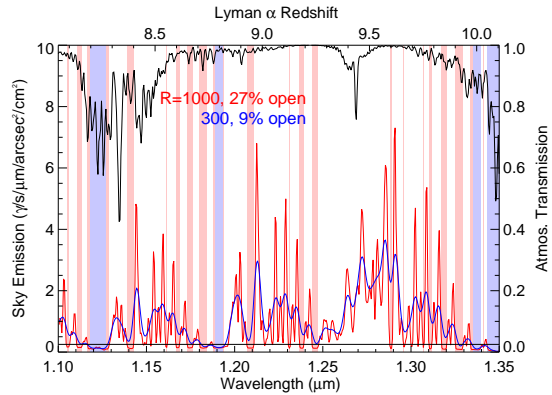}
\caption{``Windows'' in the $J$-band night sky spectrum.  The black
line indicates the transmission of the night sky ({\it scale on
right}).  We plot the night sky spectrum at two resolutions ({\it
$R=1000$, red line; $R=300$, blue line}) to indicate regions below
1/4 of the mean background (blue shading for moderate resolution, red
shading for high resolution). At moderate resolution, only a small
number of windows are available.}
\label{fig:OH}
\end{figure}

The strongest constraints on the creation and transmission of
Ly$\alpha$ come from the cosmic microwave background. The {\it WMAP}
observation of the Thompson optical depth to reionization, $\tau_{e} =
0.17 \pm 0.4$ \citep{Kog03}, constrains the
parameters in Equation~\ref{eqn:lya}.  By modeling the
reionization process,  Cen (2003a,b) and Sokasian et al. (2003b) find
that a large Thomson optical depth requires most or all
of the following at high redshift: a top-heavy IMF of extremely
low-metallicity stars, a high star formation efficiency in low-mass
halos, f$_{\rm esc} \gtrsim 0.3$, a positive tilt in the
matter power spectrum, or an additional source of ionizing photons.  Except for
escape fractions extremely close to unity, these conditions are all
conducive to observing $z > 7$ galaxies in Ly$\alpha$.

Although much uncertainty surrounds the search for
Ly$\alpha$ in $z > 7$ galaxies, we argue that the distinct possibility
of success motivates an attempt with present-day technology.  As an
illustration, we consider two specific scenarios that
are both detectable in Ly$\alpha$ by design.  They differ by a factor
of $\sim$3.4 in the flux of observable Ly$\alpha$ photons.  They are:
\begin{enumerate}
\item {\bf Plausible:} An IMF rich in massive stars with a Salpeter
slope, containing only stars in the range 50~--~500 M$_{\sun}$, with a
metallicity of Z$_{\sun}=0$, an escape fraction of f$_{\rm esc}= 0.1$,
$\tau_{\rm dust}=0$, and f$_{\rm IGM} = 0.25$, which is appropriate
for a range of IGM wind models, yielding ${\rm F_{Ly\alpha,obs}=6.4
\times 10^{42}\ erg\ s^{-1}/[M_{\sun}/yr] }$ of detectable Ly$\alpha$
\citep{Sch03,San03}.
\item {\bf Optimistic:} An extremely top-heavy IMF composed of only
300-1000 M$_{\odot}$ zero-metallicity stars \citep{Bro01}, $\tau_{\rm
dust} = 0$, ${\rm f_{esc}}=0.35$ to maximize the observable Ly$\alpha$
emission based on the fiducial \citet{San03} models, and ${\rm
f_{IGM}=1}$, appropriate for an IGM model with some ionizing
background in which strong galactic winds clear the Stromgren spheres
around galaxies and shift Ly$\alpha$ toward the red \citep{San03}.
This combination yields ${\rm F_{Ly\alpha,obs}=2.1 \times 10^{43}}$
${\rm erg\ s^{-1}/[M_{\sun}/yr] }$ of detectable Ly$\alpha$.
\end{enumerate}

Both scenarios make assumptions that are relatively favorable for
detecting Ly$\alpha$ emission, such as the presence of a top-heavy IMF
in high-redshift galaxies.  If regions of intense star formation are
sufficiently chemically enriched by $z\sim 8$ to have a Salpeter IMF
from ${\rm 1-100\ M_{\sun}}$, ${\rm F_{Ly\alpha,em}}$ is reduced by a
factor of $4-12$, depending on metallicity.  With $6.3\times 10^8$
years of cosmic time available for star formation before $z = 8$,
delaying chemical enrichment sufficiently long may require suppressing
star formation at early times, as would be the case if the power
spectrum were reduced on small scales \citep[e.g.,][]{Yos03a,Yos03b}.
In addition, inhomogeneous structure formation may lead to pockets of
low-metallicity star formation even at late epochs \citep{Sca03}.
Observationally, rest-frame equivalent widths of Ly$\alpha$ as late as
$z \sim 5.7$ support the hypothesis that the IMF may be top-heavy at
late epochs \citep[e.g.,][]{Rho03}.  More indirect arguments in favor
of massive and/or low-metallicity early stars come from comparisons of
the star formation history of the universe to the requirements for
reionization \citep[e.g.,][]{Fer02} and from fossil elemental
abundances \citep{Bea03}.

\section{Identifying and Mapping High-redshift Sources}

\begin{figure}
\plotone{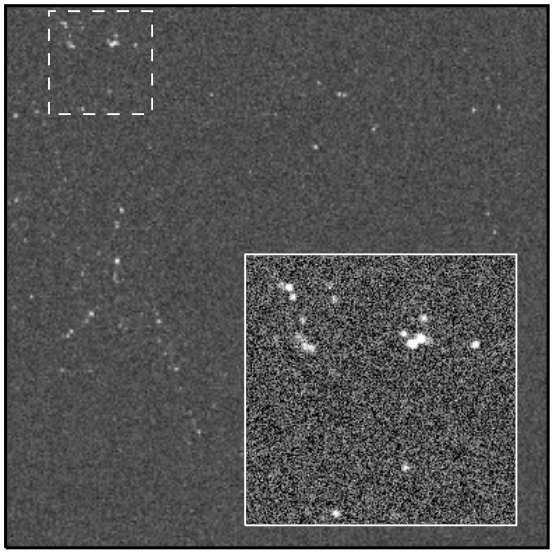}
\caption{A realization of a 32-hour $2^{\prime}\times2^{\prime}$
observation of the ``optimistic'' scenario with an 8-meter telescope.
We simulate detection of the objects in the \citet{Spr03a} Q5 ${\rm
10\ h^{-1}\ comoving\ Mpc}$ model.  The $R=125$ narrow-band filter we
assume is actually 44\% thicker in redshift depth than this simulated
observation.  There are 56 detectable sources in the field.  We show
the dotted inset enlarged for clarity.}
\label{fig:sim}
\end{figure}

A simple argument shows that for even a moderate Ly$\alpha$ emitter,
detecting Ly$\alpha$ from the ground at high spectral resolution is
easier than detecting the continuum emission.  For example, in the
J-band window at 1.122$\mu m$ (Ly$\alpha$ at z=8.2; see Sec. 3.1),
assuming a Mauna Kea sky background and observations in which sky
noise dominates, the signal-to-noise ratio of a suitably located
$R=200$ observation of Ly$\alpha$ becomes equal to or better than that
of a broad-band continuum observation ($R=4$) when the rest-frame
$\mewe \gtrsim 27$~\AA.  At $R=2000$, the threshold is $\mewe \gtrsim
8$~\AA.  Observed rest-frame Ly$\alpha$ equivalent widths of $z=5.7$
galaxies exceed 150~\AA\ \citep{Rho03} and starburst models predict
equivalent widths far in excess of 100~\AA\ for most young starburst
scenarios \citep{Bro01,Sch03}.  Thus, detection in Ly$\alpha$ is
likely the most powerful way to identify and map very high-redshift
star formation from the ground.

Observing spectral features between the strong atmospheric lines helps
to minimize the background in the near-infrared.  Although there are
existing spectrographs with high enough resolution to observe between
OH lines, the relatively low number density of detectable objects
predicted by simulations argues for blind searches for Ly$\alpha$
emitters in a large, contiguous field ($\gtrsim {\rm
few}~\square^{\prime}$).  Because narrow-band filters with $R \gtrsim
200$ are not widely available, the most effective current strategy is
to identify the broadest OH-free windows in the night sky;
Fig.~\ref{fig:OH} illustrates some windows in the $J$-band, including
a wide window at $z=8.2-8.3$ with $\sim$80\% (variable) atmospheric
transmission.\footnote{We use the Gemini sky spectrum from
http://www.gemini.edu.}

         \subsection{A Simulated Observation with an 8-meter Telescope}

We assume each star-forming particle in our simulation is a source of
Ly$\alpha$ photons, converting star formation rate to F$_{\rm
Ly\alpha,obs}$ as described in \S\ref{sec: theory}.  For an $R=125$
narrow-band filter in the $z=8.227$ $J$-band window, $0\farcs35$
seeing, a total system throughput of 20\%, and the detector parameters
appropriate for a modern detector (the NIRI detector on Gemini), we
simulate 32-hour observations of Ly$\alpha$ for the ``optimistic'' and
``plausible'' scenarios.  Fig.~\ref{fig:sim} shows a $2^{\prime}
\times 2^{\prime}$ corner of the ``optimistic'' simulation; the
thickness of the simulation is 44\% less than the thickness of an
$R=125$ observation, so the simulation actually underrepresents the
number of sources expected in this scenario.

Allowing for the depth and angular size of the simulation, the volume
probed by the entire Q5 model is 5.14 times larger than the volume of
a single $R=125$ narrow-band filter observation with a $2^{\prime}
\times 2^{\prime}$ detector.  We analyze the simulated observations
with the SExtractor software \citep{Ber96}.  For the ``optimistic''
case, we find 176 sources in the entire simulation volume to a S/N of
$\gtrsim3.4$.  Although the segment in Fig.~\ref{fig:sim} is in a
well-populated part of the simulation, with 56 detectable sources, the
detectable objects do cover the volume well in the sense that any
$2^{\prime} \times 2^{\prime}$ field would include observable
galaxies.  In the ``plausible'' simulation case, we detect 23 sources.
The objects are clustered, but a $2^{\prime} \times 2^{\prime}$
observation would reveal no sources $\lesssim$20\% of the time.

\begin{figure}
\plotone{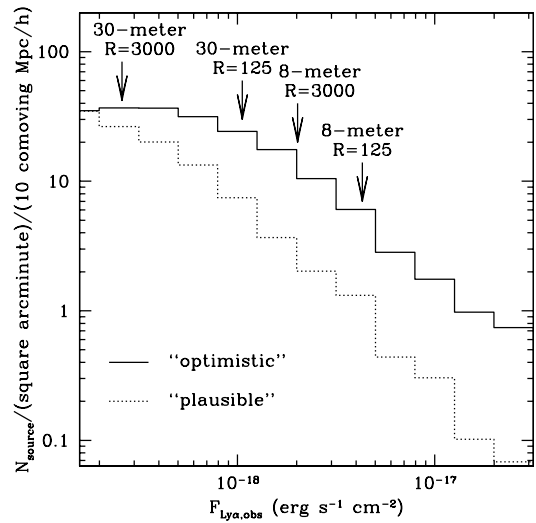}
\caption{Ly$\alpha$ fluxes in the $z=8.227$ $J$-band window.  We plot
number counts per square arcminute in the \citet{Spr03a} Q5
${\rm 10h^{-1}\ comoving\ Mpc}$ simulation, converting star formation
rates to observable Ly$\alpha$ fluxes via the ``plausible'' and
``optimistic'' scenarios outlined above.  The arrows indicate
detection limits for 32-hour exposures with 8-meter and 30-meter
telescopes at moderate ($R=125$) and high ($R=3000$) resolutions.}
\label{fig:limits}
\end{figure}

\subsection{Identifying Low-redshift Contaminants}

Although detecting galaxies via their Ly$\alpha$ emission is the
primary challenge, separating the sources from possible contaminants
is equally important.  Both continuum and emission-line sources will
appear in the narrow-band images.  Contaminants from lower redshifts
must be identified with additional data.  Deep broad-band images in
both the optical and near-infrared will serve this purpose.  For
example, for a 32-hour narrow-band observation at $R=125$, $J$-band
data that is deeper than $\sim$25.9 (AB) should reveal all continuum
sources in the narrow-band image.  Many upcoming imaging surveys will
exceed this sensitivity.

Extremely high redshift galaxies may be difficult to detect in the
continuum, even with HST.  The brightest galaxy known at $z\sim6$
\citep{Dic03} would be detectable at $z=8.2$, without evolution, in
the upcoming HST/NICMOS ultra-deep field observations.  But at $z \sim
8$ galaxies should have lower star formation rates and less stellar
mass.  Even if the continuum flux is detectable, the target objects
will be identifiable as $z$-band dropouts and distinguishable from
lower redshift contaminants which will appear in the optical bands.

The remaining possible contaminants are low-$z$ objects with strong
emission lines and no detectable continuum.  In all cases, however,
the emission lines must be prohibitively strong for the objects to
remain undetected in extremely deep broad-band images.  For the
scenario described above, if a detected emission line in the
1.122~$\mu$m window is actually [O\,{\sc II}]($\lambda3727$), the $z=2.01$
galaxy would have to have an unphysical equivalent width of
$\sim$800~\AA\ or more (rest-frame) to remain undetected in a deep
F606W observation (29.3 magnitude limit at $3\sigma$, corresponding to
the HDF-North data).  Other emission lines place similar requirements
on the high-redshift objects, including both QSO lines and strong
galaxy lines.  Thus, the sources that appear in the narrow-band image
but are non-detections in deep optical continuum images are almost
certainly $z > 7$ galaxies.

\section{Goals for Present and Future Facilities}  

Many far-reaching science goals motivate devoting an enormous effort
to the study of extremely high-redshift ``first-light'' objects.  Some
of these goals are: (1) measuring the luminosity function of $z>7$
galaxies, to compare with the number of ionizing photons needed to
reionize the universe, and to measure the global star formation rate
at these redshifts, (2) mapping high-redshift sources and measuring
their kinematics, to observe their clustering properties and compare
with models of early galaxy formation, (3) quantifying high-$z$
Stromgren spheres in the IGM from Ly$\alpha$ line profiles, and (4)
directly probing the IMF and metallicity of early stars with, e.g.,
He\,{\sc II}($\lambda1640$) and, eventually, other spectral lines.

Accomplishing these goals will require a large, complementary set of
observational facilities.  The {\it James Webb Space Telescope}, with
broad-band sensitivities of $\sim$2 nJy, will detect early stars in
their continuum light.  Narrow-band imaging will likely remain the
strongest tool of ground-based facilities in the near-infrared.
Fig.~\ref{fig:limits} shows the expected performance of 8-meter and
future 30-meter ground-based telescopes.  Assuming modern detector
characteristics, we plot the fluxes from star-forming regions at
$z=8.227$.  Clearly, 8-meter class telescopes have great near-term
potential for advancing the study of early star formation.  Narrower
filters ($R \geq 1000$) would improve sensitivity substantially,
although they would require some development in both filter and
detector technology. \footnote{This technology is one goal of the
DAZLE spectrograph, under development for the VLT
(http://www.aao.gov.au/dazle/).}$^,$
\footnote{At, e.g., $R=3000$, a long NIRI observation would actually
be detector-noise dominated; thus, the development of low read noise
and low dark current detectors is another important component in the
search for extremely high-redshift objects.}  Looking further into the
future, the detailed physical study of these objects --- an
exploration down the luminosity function, measurements of Ly$\alpha$
line profiles, and the detection of fainter spectral emission lines
--- is probably the role of the next generation of large ground-based
telescopes.  The timely discovery of high-z Ly$\alpha$ emitters
is crucial to guide the design of future facilities,  since probing the epoch
of reionization is one of their central scientific goals.

In this Letter, we motivate the search for high-redshift galaxies by
presenting plausible scenarios in which the Ly$\alpha$ emission from
$z \approx 8$ galaxies is observable with 8-meter-class telescopes.
The discovery of $z > 7$ objects is a crucial component of the quest
to understand the evolution of the early universe.  They existed at an
epoch when the universe was still partly neutral, and they are the
best candidates for supplying the radiation that reionized the
universe.  Although we are able to combine many areas of recent
research to arrive at an educated guess about the properties of $z >
7$ objects, there are still many critical unanswered questions about
the nature of reionization.  With the discovery and study of $z > 7$
objects, many of these questions will be answered.  We stand on the
verge of probing reionization directly.

\acknowledgments We thank Margaret Geller for encouraging us to write
this paper, and for reading a draft, and we thank Rob Kennicutt and
Steve Furlanetto for insightful comments.  Support for EJB and RAD was
provided by NASA through Hubble Fellowship grants \#HST-HF-01135.01
and \#HST-HF-0128.01-A, respectively, awarded by the Space Telescope
Science Institute, which is operated by the Association of
Universities for Research in Astronomy, Inc., for NASA, under contract
NAS 5-26555.

\end{document}